\documentclass{article}
\usepackage{algorithm}
\usepackage{algorithmic}
\usepackage{epstopdf}
\usepackage{verbatim}
\usepackage{amsmath}
\usepackage{epsfig}

\begin{document}

\title{A Local Approach for Identifying Clusters in Networks.}
\author{
Sumit Singh \footnote{Supported by MHRD, Govt. of India.  PS: The work was under-progress but is stopped due to some reasons.} \\
Indian Institute of Technology Kanpur,\\
       Kanpur, India.\\
       singhsumitbit@gmail.com
 }
 \date{}
\maketitle
\begin{abstract}
Graph clustering is a fundamental problem that has been extensively studied both in theory and practice. The problem has been defined in several ways in the literature and most of them have been proven to be NP-Hard. Due to their high practical relevancy, several heuristics for graph clustering have been introduced which constitute a central tool for coping with NP-completeness, and are used in applications of clustering ranging from computer vision, to data analysis, to learning. There exist many methodologies for this problem, however most of them are global in nature and are unlikely to scale well for very large networks.
In this paper, we propose two scalable local approaches for identifying the clusters in any network. We further extend one of these approaches for discovering the overlapping clusters in these networks. Some experimentation results obtained for the proposed approaches are also presented.
\end{abstract}

\section{Introduction}
Identifying clusters in complex networks is difficult and has attracted a lot of attention in recent years as well as in the past, especially in computer science community, statistical physics community \cite{newman2, mendes}, biological community \cite{wagner} and applied mathematics community.  Popular examples include the food webs, metabolic networks \cite{barabasi,wagner}, protein interactions \cite{jruan}, Internet and world wide web \cite{Broder:2000:GSW:346241.346290,Faloutsos:1999:PRI:316194.316229}, communication and distributed networks, and the social networks \cite{zachary,mislove-2009-socialnetworksthesis}. To extract clusters in such networks, one typically chooses an objective function that captures the above intuition of a cluster/community as a set of nodes with better internal connectivity than external connectivity. Generally, the objective is typically NP-hard to optimize exactly \cite{Leighton:1999:MMM:331524.331526,newman}, one employs heuristics \cite{newman,dhillon} or approximation algorithms \cite{Arora:2009:EFG:1502793.1502794} to find sets of nodes that approximately optimize the objective function and that can be understood or interpreted as 'real' communities. \\

In this work, we have explored random walk based techniques and tried to improve their performance by some heuristics which are very popular in statistical physics community. The heuristics seems to help algorithm converge faster than usual on several standard datasets containing upto $30$ million nodes. 

\subsection{Related Work}
The earliest attempts dating back to the seventies were based on local search, with the Kernighan-Lin heuristic \cite{kernighan} achieving the best results in this class. Spectral methods \cite{Hendrickson:1995:ISG:203046.203060, Karypis:1997:MHP:266021.266273, Pothen:1990:PSM:84514.84521}, which found widespread use in the eighties, compute eigenvectors, which are embeddings of the graph onto the real line. In the early nineties, the linear programming methods were shown to find better cuts \cite{Lang:1993:FNC:313559.313755} than Kernighan-Lin and the spectral method but were too slow to be used. In the mid-nineties, a multilevel heuristic embodied in the package METIS was introduced \cite{Karypis:1998:FHQ:305219.305248}. This method is very fast and produces cuts that are almost as good as those obtained by linear programming based methods in practice.

An optimization version of clustering, \textit{Sparsest Cut Problem} has been extensively studied. The Sparsest cut problem is to bipartition the vertices so as to minimize the ratio of the number of edges across the cut divided by the number of vertices in the smaller half of the partition. This problem has been proven to be NP-Complete \cite{Leighton:1999:MMM:331524.331526, DBLP:conf/sofsem/2006}. Leighton and Rao \cite{Leighton:1999:MMM:331524.331526} used linear programming to obtain $O(\log n)$ approximation of the sparsest cut. Arora, Rao and Vazirani \cite{Arora:2009:EFG:1502793.1502794} improved this to $O(\sqrt{\log n})$ through semi-definite programming. Faster algorithms obtaining similar guarantees have been constructed by Arora, Hazan and Kale \cite{Arora:2004:NA:1032645.1033179}, Khandekar, Rao and Vazirani \cite{Khandekar:2009:GPU:1538902.1538903}, Arora and Kale \cite{Arora:2007:CPA:1250790.1250823}, and Orecchia, Schulman, Vazirani, and Vishnoi \cite{Orecchia:2008:PGV:1374376.1374442}. Other related problems which have been extensively studied by theoreticians are \textit{Balanced Min-cut} problem and \textit{Densest Subgraph} problem.

In early 2000, some min-cut based techniques were suggested that essentially iteratively finds the min-cut in the graph and removes it to form partitions \cite{Flake:2002:SIW:619073.621934, erez}.
Few years back, modularity based approach suggested by Newman \cite{newman} from statistical physics community gained a lot of popularity mainly because of being practical and being able to quantify the goodness of a cluster. Later it was shown that maximizing modularity is an NP-Hard problem \cite{DBLP:journals/tkde/BrandesDGGHNW08} however the naive greedy approach has been shown to work well in practice. 

In many networks such as social networks or author collaborative network, one expects a node to be in more than cluster. Hence we further look onto to unveil this overlapping structure in the networks. One of the first attempts to obtain the overlapping community structure of a graph appears in \cite{lancichinetti-2009-11, palia}. The approach in \cite{palia} is based on retrieving all cliques of the graph; however, this operation turns out to be prohibitive for large graphs. A more efficient algorithm is given in \cite{lancichinetti-2009-11}, which finds communities by maximizing a certain fitness function. Recently, a very comprehensive study is carried out by Kovács, Palotai, Szalay and Csermely $\cite{csermely}$ where they propose and evaluate the efficiencies of several methods that aim to discover the overlapping community structure in both directed and undirected graphs. 
\subsection{Local Clustering}
A graph algorithm is a local algorithm if given a particular vertex as input, at each step only examines the vertices connected to those it has seen before. The use of a local algorithm naturally leads to the question of \textit{in which order one should explore the vertices of a graph}. While it may be natural to explore vertices in order of shortest-path distance from the input vertex, such an ordering is a poor choice in graphs of low-diameter, such as social network graphs \cite{Leskovec:2008:PVL:1367497.1367620}. A natural inclination is then to choose neighbours randomly or the vertices which come first in the random walk. Our proposed approaches are based on this inclination which has been used for clustering in the past \cite{dongen}. 

Our choice of choosing vertices randomly or considering probabilty distributions is further strengthened by the following results. Earlier, Meila and Shi \cite{DBLP:conf/nips/2000,Meila01arandom} had shown that the normalized cut of a graph, one of the variants of graph conductance, can be expressed in terms of the transition probabilities and the stationary distribution of a random walk in the graph, thus linking the mathematics of random walks to those of cut-based clustering. Orponen and Schaeffer \cite{DBLP:journals/corr/abs-0810-4061} in turn expressed the absorption times of a random walk in a graph in terms of the eigenvectors of the graph's Laplacian and use their locally computable approximation of the Fiedler vector \cite{DBLP:conf/wea/2005} to obtain an approximation of the absorption times. This links the random walks to spectral clustering, which relates to cut-based methods. Recently, Teng \cite{DBLP:conf/tamc/2010} introduced a \textit{Laplacian Paradigm} for designing nearly linear time and scalable algorithms which discusses a lot of theoretical aspects of these approaches.

\section{Problem Definition}
Let $G=(V,E)$ be any undirected network, then a cluster $S$ of $G$ is a subset of $V$ that is richly intra-connected but sparsely connected with the rest of the graph. The quality of a cluster can be measured by conductance which is defined as the ratio of the number of external connections to the total number of connections. 

Let $E(S,V-S)$ be the set of edges where an edge $e=(u,v)$ belongs to this set if $u \in S$ and $v \in V-S$ or vice versa. Let $d_i$ be the degree of $i^{th}$ vertex, then the volume $\mu(S)$ of any set $S \subseteq V$ is defined as sum of degrees of every vertex in $S$. The conductance is formally defined as : \\

\begin{eqnarray*} 
 \alpha(S)= \frac{|E(S,V-S)|}{min \{ \mu(S),\mu(V-S) \} } \\
\end{eqnarray*} 
The conductance of the graph $G$ is defined as the minimum conductance over the conductance of all subsets $S$ of $V$. The subset $S$ defines the cut and the partitions defined are $S$ and $V-S$. Clustering is thus presented as an optimization problem : Given an undirected graph $G$ and a conductance parameter, find a cluster $C$ such that $\Phi(C) \leq \phi$, or determine no such cluster exists.

\textit{Graph Preprocessing} \\
We have introduced self loops in the transition probability of the given graph $G$ and the graph is therefore assumed to be ergodic. A graph is said to be ergodic if aperiodic and irreducible. Introducing self loops ensures aperiodicity and the fact that graph is undirected and connected ensures that it is irreducible.

\section{Proposed Approach} 
The paper suggests two approaches for the clustering problem and extends one of these approaches to discover overlapping clusters. The underlying observation behind these approaches is that a random walk started at some vertex $v$ will initially be trapped in the cluster to which $v$ belongs. Thus carrying out this process again and again for walks of some length, we will revisit the nodes in the cluster of $v$ more frequently. One of the primary motivation of such random walk based algorithms comes from the fact that was implicit in the analysis of the volume estimation algorithm of Lovasz and Simonovits \cite{DBLP:journals/rsa/LovaszS93} that one can find a cut with small conductance from the distributions of the steps of the random walk starting at any vertex from which the walk does not mix rapidly. However when the algorithm based on above intuitions were carried by us on various networks, it was observed that a large number of iterations are required to find good clusters and to ensure convergence. We therefore extend this naive approach so as to give good results in less time.

\subsection{Distribution Based Approach}
Informally, conductance of any set $S$ defines the probability of going out of set $S$ given that we are in set $S$. Since we do not know the cluster $S$ apriori, we try to estimate set $S$ by trapping ourself heuristicly within some set $X$. The set $X$ can be thought of as an estimation of $S$ that is likely to improve as the algorithm progresses. The approach considers a probability distribution vector which is initially concentrated on the vertex $v$ whose cluster we are trying to find. This distribution is then evolved based on the transition probabilities of the network. The transition probability from any node $x$ to any node $y$ are strongly aperiodic and is defined as follows for any network:

\[
  P[x \rightarrow y] = \left\{
  \begin{array}{l l}
    \frac{1}{2} & \quad \text{if } x=y\\
    \frac{1}{2d_x} & \quad \text{if $y$ is a neighbour of $x$} \\
    0 & \quad \text{otherwise} \\
  \end{array} \right.
\]

For the exposition of the ideas the approach is formally given below in a more readable manner without ignoring the implementative issues.

\begin{algorithm}
\caption{Distribution\_Based\_Approach(vertex v)}
\label{Algorithm}
\begin{algorithmic}[1]
\STATE  Construct the probability matrix P.\\
\STATE Initialize distribution[] vector to 0 except at index v which is set to 1.\\
\STATE while(Condition) do\\
 \textrm{a) For each vertex w such that distribution[w]>0, do\\}
 \quad \textrm{i) For each vertex u neighbouring to w, increase distribution[u] by $\frac{1}{2d_w}$\\}
 \textrm{b) Truncate(distribution,v,$\alpha$).}\\
\end{algorithmic}
\end{algorithm}

\begin{algorithm}
\caption{Truncate(distribution[],vertex v,parameter $\alpha$)}
\begin{algorithmic}[1]
\STATE tmp=0;\\
\STATE For each vertex w\\ 
	if distribution[w]<$\alpha$*distribution[v]\\
	\textrm{a) tmp+=distribution[w].}\\
	\textrm{b) distribution[w]=0.}\\
\STATE ditribuition[v]+=tmp.	
\end{algorithmic}
\end{algorithm}

The initialization step in the algorithm is basically capturing the fact that the random walk starts at vertex $v$ and as the walk progresses the probability of walker to be at different states is captured in distribution vector. Note that in each step a truncation step is performed. In absence of truncation, the distribution vector will converge to stationary distribution vector of the network in a long run.  The truncation ensures two things: one is that the experiment remains local i.e. the random walk is allowed to diffuse in the network in a constrained manner and the other is that instead of starting a new walk as was done in previous approaches, we reinforce the elements that seems going out of cluster/community back into it. Thus this reduces the number and length of run(s) of the experiments. The parameters in truncation plays a dominating role for controlling  the cluster size of the network. If the parameter is small then vertices having less belonging to that cluster will also be included in the cluster. We also define a belongingness measure here for each vertex $v$, which is the ratio of distribution of the node $v$ to the distribution of the node which is presently the centre of cluster. Note that since the probability distribution is changing abruptly due to truncation step, the process described above doesn't seems to be a associated with a single/fixed random walk. However, we observe experimently that the values of distribution vector converge for a fixed value of parameter $\alpha$.

\subsection{Adaptive Walk Based Approach}
This approach is inspired by Wang-Landau algorithm \cite{PhysRevLett.86.2050} from statistical physics community. The problem that we faced in random walk based naive approach is that the walk may escape through the cluster and so we need to perform it again from starting. These things are bound to happen in ergodic networks as any markov chain tries to mix such that the corresponding distribution acheives the stationary distribution. Therefore for a network having small or moderate mixing time, the naive approach may require high number of iterations. We observe that if we increase the incoming probability for those vertices which we have visited earlier then we may be able to trap the random walk into some local minima which would ideally be the cluster itself. Therefore the task is to redefine the probabilities at certain interval of walk, so that the random walk is trapped inside the cluster of the starting vertex. The Wang-Landau algorithm in some sense does the reverse of this by defining transition probabilites such that the random walk does not trap in local minima. For theoretical reasons, the Wang-Landau algorithm is not a markov process but has been adapted to a markov process recently. There are some converging issues with this algorithm for various networks which has been resolved in the physics community \cite{PhysRevE.78.046705}. We have not yet studied the implication of similar modifications theoretically in our approach and is an interesting problem to look at. 

In the adapted approach, each vertex has an associated energy value. The probability of transition from node $u$ to node $v$ is defined as: 
\begin{eqnarray*}
P(u \rightarrow v)=min\{\frac{energy[v]}{energy[u]},1\} \\
\end{eqnarray*}

The new adapted approach is described as follows: \\

\begin{algorithm}
\caption{Adaptive\_Walk\_Based\_Approach(vertex v, parameter f, parameter $\alpha$, parameter $\beta$)}
\begin{algorithmic}[1]
\STATE For each vertex w: energy[w]=$\frac{\alpha}{d_v}$.
\STATE energy[v]=$\frac{\beta_v}{d_v}$.
\STATE current\_vertex=v.
\STATE while(Condition) do \\
		\textrm{a)} next\_vertex=random\_walk(current\_vertex,P).\\
		\textrm{b)} energy[current\_vertex]=energy[current\_vertex]*f.\\
		\textrm{c)} current\_vertex=next\_vertex.	
\end{algorithmic}
\end{algorithm}

The random\_walk function above chooses the next vertex randomly based on the current vertex and the transition probability. The value of $f$ in the Wang Landau algorithm was kept equal to $e=2.71828..$, and was then slowly decreased with further iterations. In the experiments, we performed we started with a low value of $f$ around $1.3$ and then increased it to higher values. One natural question that emerges out of this is when should the value of $f$ be changed. Though there is no correct answer for this, it depends on the size of community we want to find. The algorithm is actually defining some loose boundary for lower values of $f$ then it is filtering again with higher values of $f$. After readjusting $f$, the current vertex is reset to the original vertex for which the algorithm was called. The values of parameter $\alpha$ and $\beta$ is chosen low(small constant) and high respectively. There are still some convergence issues in this approach for any arbitary network(s) both experimentally and theoretically. 
  
\section{Overlapping Clusters}
A node in a network may belong to more than one cluster in a network. These overlapping structures can easily be observed in our social lifes. Thus a natural extension to the clustering problem is to find the overlapping clusters in a network. We used \textit{Fuzzy C Means Clustering} to resolve the problem and obtained good results. 
Fuzzy c-means is a method of clustering which allows one piece of data to belong to two or more clusters. This method (developed by Dunn in 1973 \cite{dunn} and improved by Bezdek in 1981 \cite{Bezdek:1981:PRF:539444}) is frequently used in pattern recognition. It is based on minimization of the following objective function: 

\begin{eqnarray*}
F_m=\Sigma_{i=1}^{n}\Sigma_{j=1}^{k} u_{ij}^{m} ||x_i -c_j||^2 \\
\end{eqnarray*}

over variables $u_{ij}$ and $c$ with $\Sigma_{j}u_{ij}=1$. The degree of fuzzification is controlled by parameter $m \in [1, \infty)$. $u_{ij}$ is the degree of membership of $x_i$ in the cluster $j$. $x_i$ is the $i^{th}$ $d$-dimensional measured data point. $c_j$ is the dimension centre of the cluster $j$. $||*||$ denotes the similarity between any measured data and the center. 

Fuzzy partitioning is carried out through an iterative optimization of the objective function. The first approach suggested in this paper defines distribution of each vertex $u$ when any vertex $v$ is chosen as its center. Several distribution value(s) are obtained for each node by varying the centres. These values are then treated as vector $x$ which is required in the objective function mentioned above. the cluster centre $c$ is also well defined in the above process. We then use $l_2$ norm as the similarity measure for our purpose. The degree of fuzzification was varied through some values till it ended in a local minima for the objective function defined above. The number of dimensions of vector $x$ can be varied here and for small networks, the best one of them can be chosen as a solution based on some objective function.

\section{Experiments}
In this section, we are not referring to the overlapping clustering scheme unless specificly mentioned. 
There are numerous evaluation creterion analogous to different objective functions which have been used in literature. When we talk of evaluation of any clustering algorithm,  the following two issues are expected to be addressed:
\begin{enumerate}
\item How good is the quality of the clusters?
\item How quickly can these clusters be computed?
\end{enumerate}
An inherent question in second issue is how much parallelizable/distributive/scalable is the algorithm?

\subsection{Quality of Clusters}
We use modularity to measure the goodness of the cluster(s). Given a graph $G=(V,E)$, where $m=|E|$, let $S \subseteq V$ and the number of edges whose both end-points are in $S$ be $m_S$. We define the modularity $f(S)$ of a set $S$ as $\frac{1}{4m_S}(m-E(m_S))$, where $E(m_S)$ is the expected number of edges in the random graph with same node degree sequence such that edges have both end-points in $S$.

We carried out experiments on some of the existing datasets to give an idea of how these approaches score against one of the best existing approach due to Newman. We carried out experiments on various datasets and have listed some of these in table \ref{table-mod}. The data sets are not very large datasets because the greedy algorithm based on maximizing modularity is an $O(n^2 \log n)$ time algorithm for $sparse$ graphs. It is to be noted that there are some controlling parameters in the approaches suggested by us. These parameters become important when we are not sure of community sizes. Also in approach $2$ there are some scaling factor desicions to be made whose timings we are ignorant of. This turns out to be the bottleneck problem for the performance of the approach 2.
\begin{table}
\centering
\caption{Comparision}
\label{table-mod}
\begin{tabular}{c|c|c|c|c|} \hline
Network Name & Size &Newman&Approach 1&Approach 2 \\ \hline
karate club & 34 & 0.42 & 0.42 & 0.42 \\ \hline
jazz musicians & 198 & 0.44 & 0.43 & 0.40\\ \hline
email & 1133 & 0.57 & 0.51 & 0.46\\ \hline
physicist & 27519 & 0.72 & 0.68 & 0.39\\ \hline
\hline\end{tabular}
\end{table}

\subsection{Time Required to form Clusters}
The datasets choosen for these experiments are from \cite{mislove-2007-socialnetworks}. The results presented in this section are for $orkut$ named dataset which has over $3$ million nodes and around $118$ million edges. All the results, we present have been obtained from a single core processor.  The time required by the adaptive random walk based approach has not been discussed due to the uncertained parameters in this approach. This is because though the method is very local but it's convergence is still a issue for massive networks. 

\begin{figure}
\epsfig{file=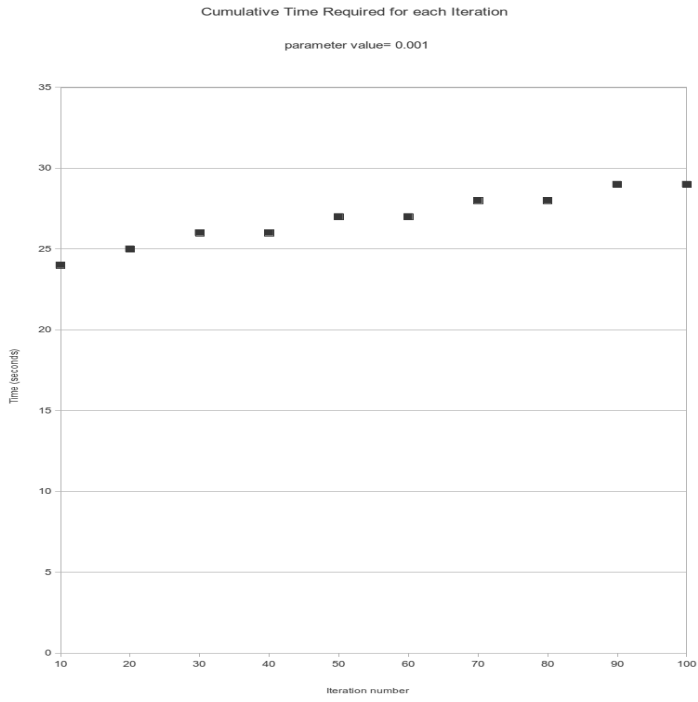,height=4in, width=4in}
\caption{Time vs Iterations(parameter value=$10^{-3}$)}
\label{f1}
\end{figure}

\begin{figure}
\epsfig{file=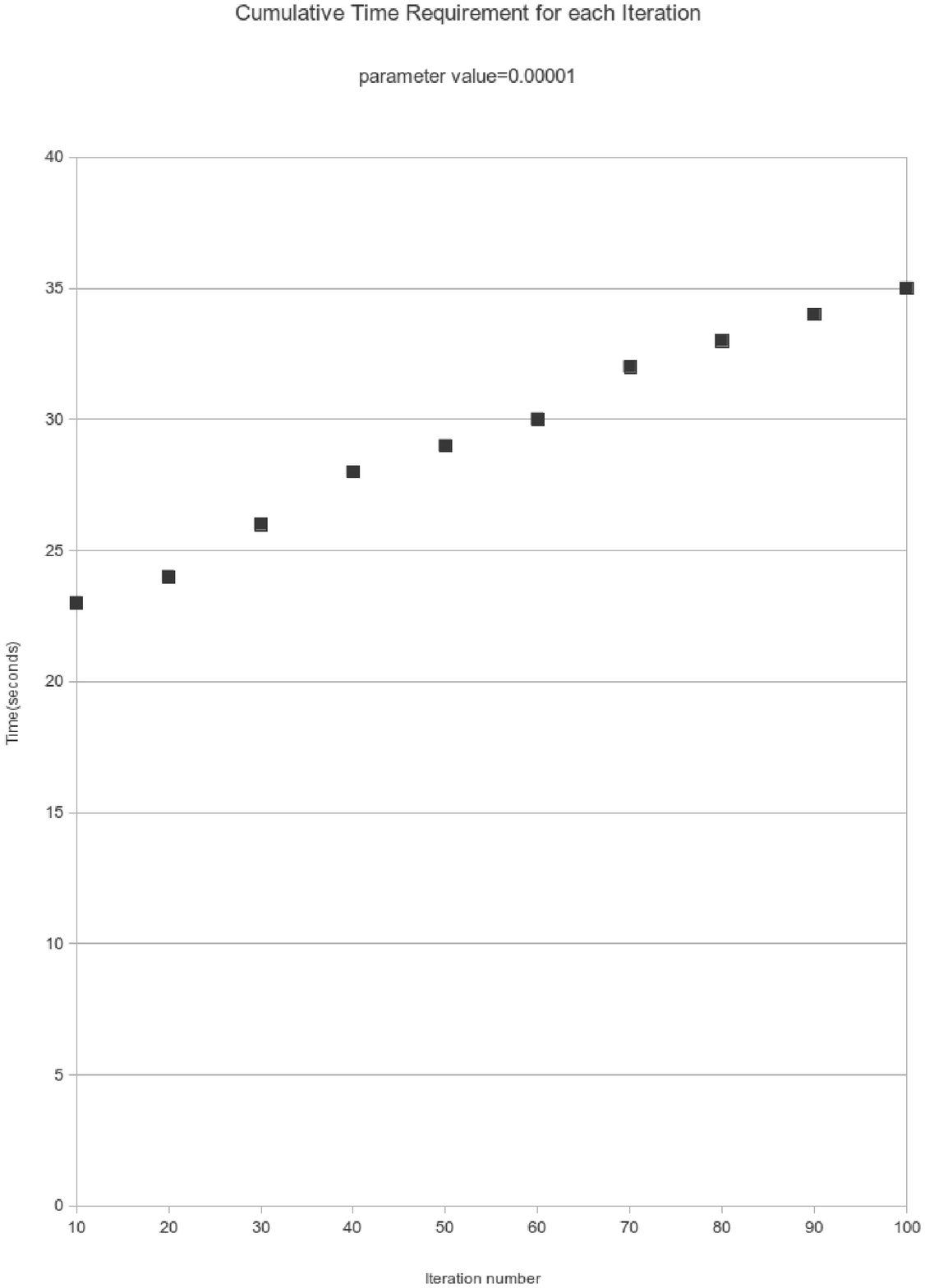,height=4in, width=4in}
\caption{Time vs Iterations(parameter value=$10^{-5}$)}
\label{f2}
\end{figure}

\begin{figure}
\epsfig{file=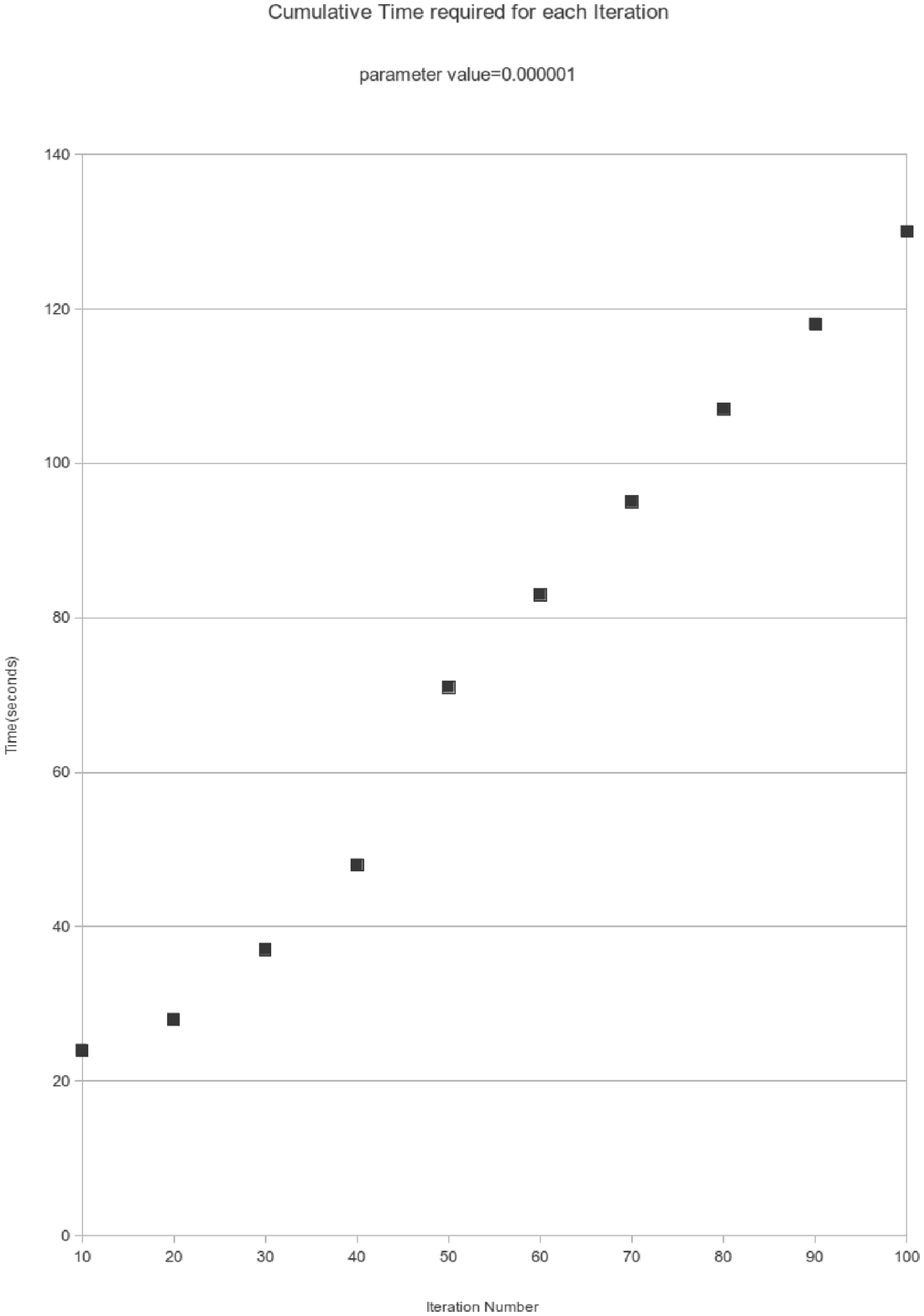,height=4in, width=4in}
\caption{Time vs Iterations(parameter value=$10^{-6}$)}
\label{f3}
\end{figure}

\begin{figure}
\epsfig{file=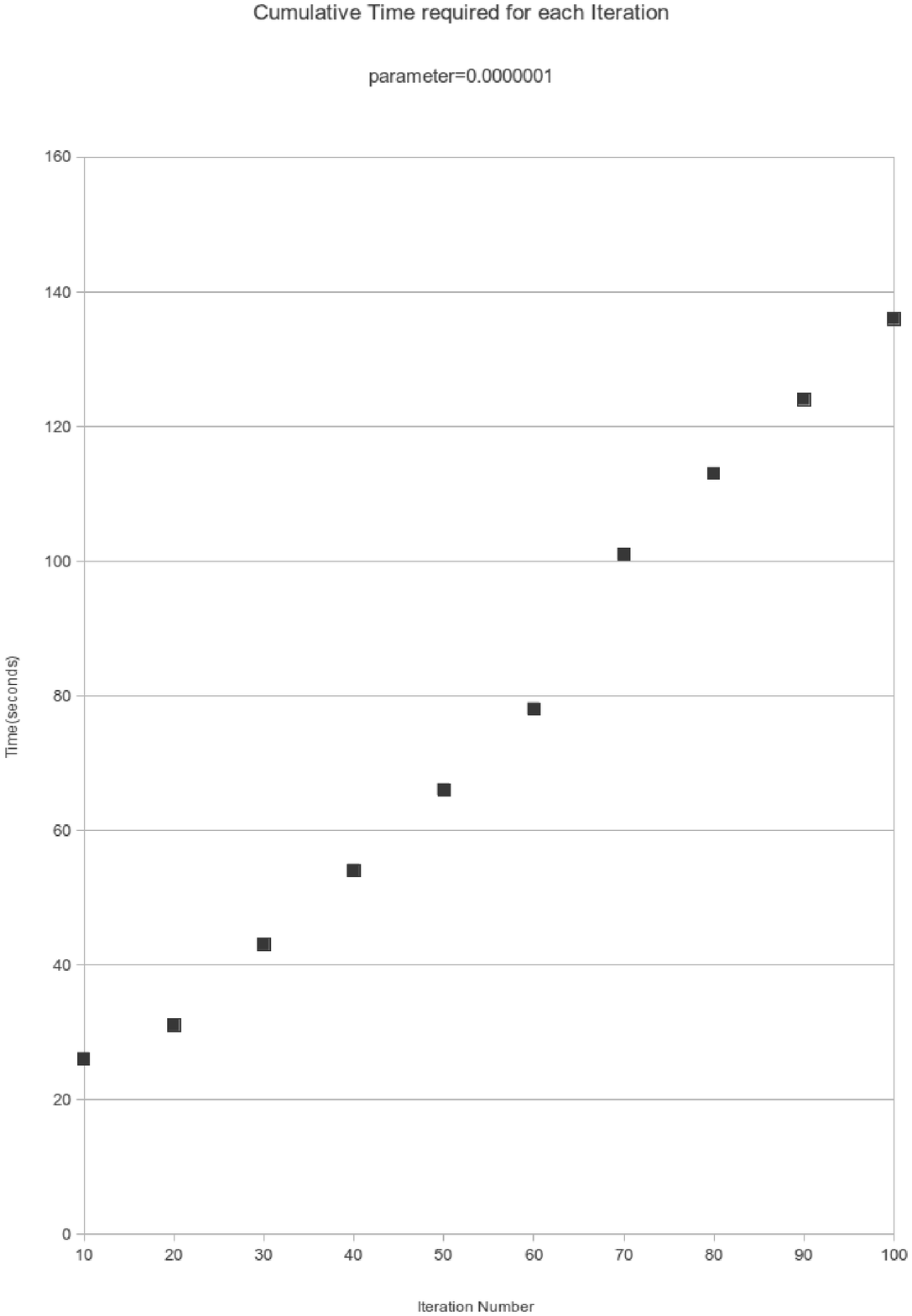,height=4in, width=4in}
\caption{Time vs Iterations(parameter value=$10^{-7}$)}
\label{f4}
\end{figure}

\begin{figure}
\epsfig{file=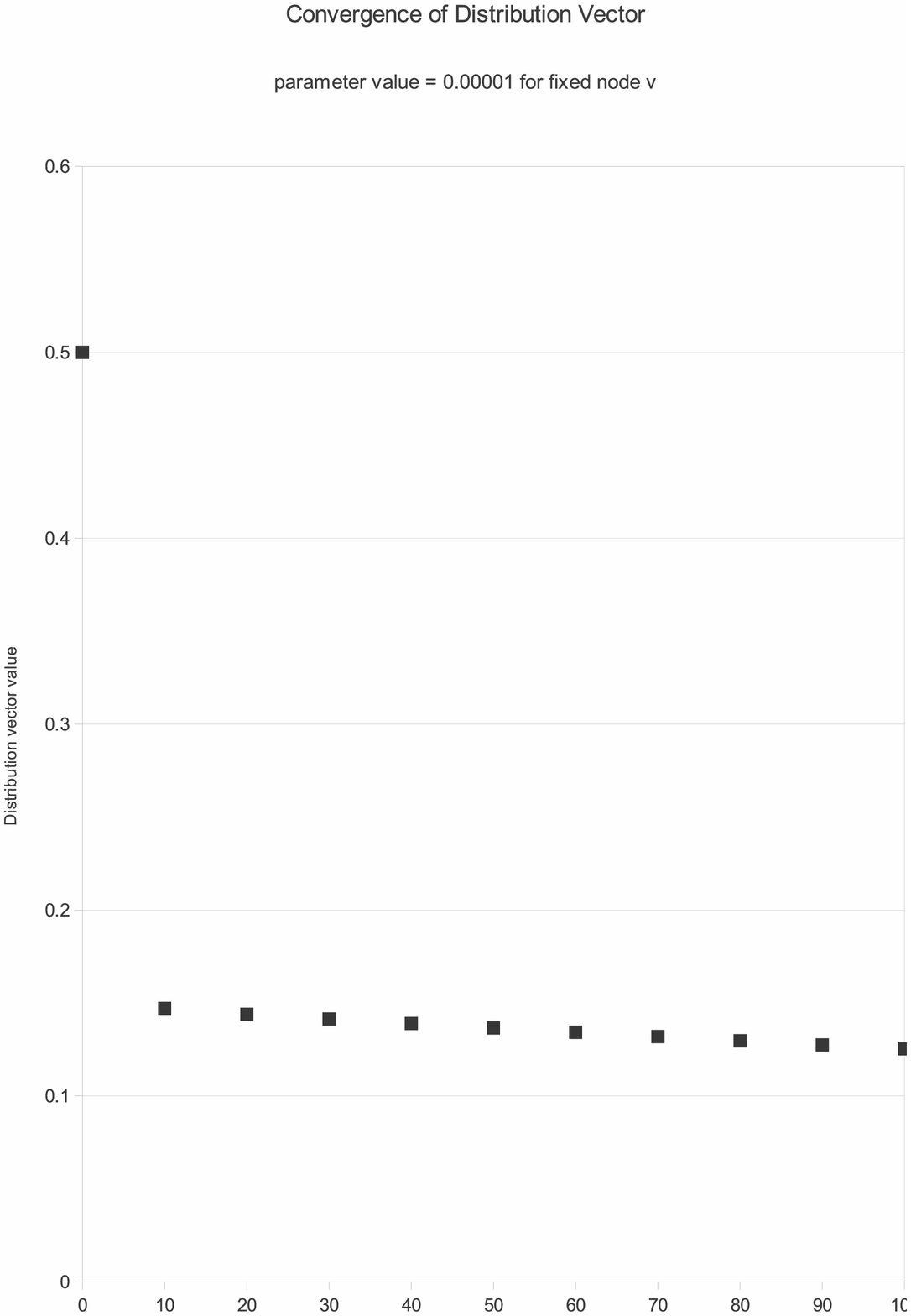,height=4in, width=4in}
\caption{Convergence vs Iterations(parameter value=$10^{-5}$)}
\label{f5}
\end{figure}

\begin{figure}
\epsfig{file=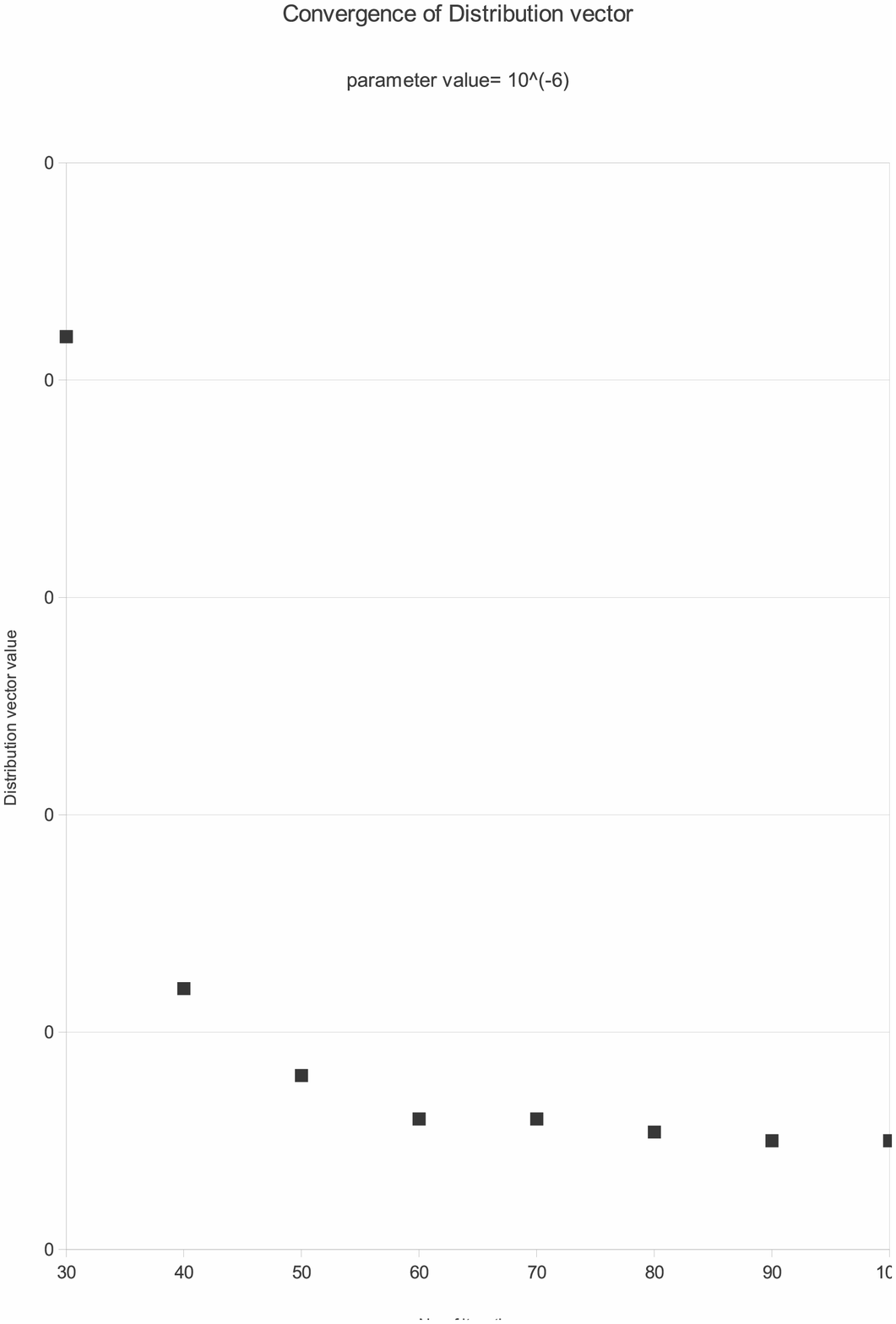,height=4in, width=4in}
\caption{Convergence vs Iterations(parameter value=$10^{-7}$)}
\label{f6}
\end{figure}

The timing results for the distribution based approach is shown in figures \ref{f1}, \ref{f2}, \ref{f3}, \ref{f4} followed by the some examples of convergence in figures \ref{f5} , \ref{f6}. All the figures \ref{f1}, \ref{f2}, \ref{f3}, \ref{f4} show the average relation between the number of iterations and the time taken. The first iteration usually requires around $24$ seconds due to file read operation. It is to be noted that the computation time increases with lower parameter value but later stablize.  
Also if the community size is less than $2000$, then the computational time required for finding one cluster is around 10 seconds. Thus it take few hours at most to naively compute all the clusters. Using architects like $CUDA$ or other parallel environments, we expect the algorithm to terminate within some minutes.

There is another and more important issue of {\bf convergence} of these approaches. We note that for very large networks, the distribution based scheme does seems to converge well however the same is not true for the adaptive walk based scheme mainly due to the reason of being heavily parametrized. The figures \ref{f5},\ref{f6} show the convergence of values in distribution vector for a particular node in distribution based approach.
\subsection{Overlapping Clusters}
\begin{figure}
\centering
\epsfig{file=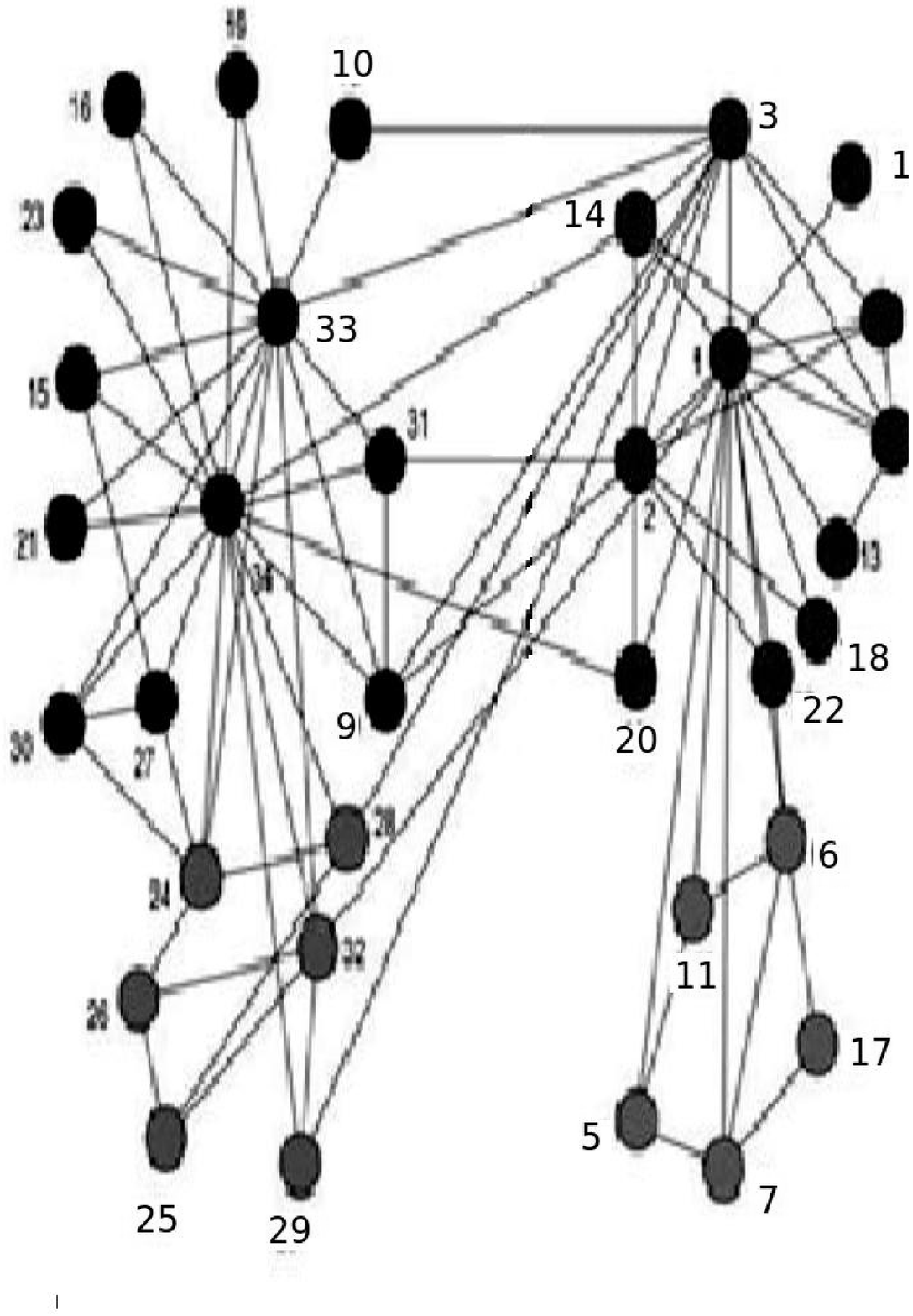,height=3in, width=3in}
\caption{Zachary Network}
\label{f0}
\end{figure}

We consider the famous network from the social science literature, the 'karate club' of Zachary \cite{zachary} (figure \ref{f0}). The network is of particular interest because, shortly after the observation and construction of the network,
the club in question split in two as a result of an internal dispute. We apply the fuzzy c-means based algorithm and discover a third overlapping community apart from these two which are mostly the vertices lying on boundaries of the two original sets(will update soon with coloured pictures).
  
\section{Conclusion}
We have proposed two approaches for the graph clustering problem in undirected graphs. There are still some improvements that can be expected in adaptive walk based approach which might improve it's convergence.  We also suggested an  approach for finding overlapping clusters in these graphs. It would be interesting to see how these approaches with appropriate modications can applied to directed graphs. 
\bibliographystyle{abbrv}
\bibliography{sigproc}  

\begin{thebibliography}{10}

\bibitem{Arora:2004:NA:1032645.1033179}
S.~Arora, E.~Hazan, and S.~Kale.
\newblock $0(\sqrt {\log n)}$ approximation to sparsest cut in $0(n^2)$ time.
\newblock In {\em Proceedings of the 45th Annual IEEE Symposium on Foundations
  of Computer Science}, pages 238--247, Washington, DC, USA, 2004. IEEE
  Computer Society.

\bibitem{Arora:2007:CPA:1250790.1250823}
S.~Arora and S.~Kale.
\newblock A combinatorial, primal-dual approach to semidefinite programs.
\newblock In {\em Proceedings of the thirty-ninth annual ACM symposium on
  Theory of computing}, STOC '07, pages 227--236, New York, NY, USA, 2007. ACM.

\bibitem{Arora:2009:EFG:1502793.1502794}
S.~Arora, S.~Rao, and U.~Vazirani.
\newblock Expander flows, geometric embeddings and graph partitioning.
\newblock {\em J. ACM}, 56:5:1--5:37, April 2009.

\bibitem{barabasi}
A.~Barabasi, B.~Tombor, H.~Jeong, R.~Albert, and Z.~Oltvai.
\newblock The large-scale organization of metabolic networks.
\newblock {\em Nature}, 407:651--654, 2000.

\bibitem{Bezdek:1981:PRF:539444}
J.~C. Bezdek.
\newblock {\em Pattern Recognition with Fuzzy Objective Function Algorithms}.
\newblock Kluwer Academic Publishers, Norwell, MA, USA, 1981.

\bibitem{DBLP:journals/tkde/BrandesDGGHNW08}
U.~Brandes, D.~Delling, M.~Gaertler, R.~G{\"o}rke, M.~Hoefer, Z.~Nikoloski, and
  D.~Wagner.
\newblock On modularity clustering.
\newblock {\em IEEE Trans. Knowl. Data Eng.}, 20(2):172--188, 2008.

\bibitem{Broder:2000:GSW:346241.346290}
A.~Broder, R.~Kumar, F.~Maghoul, P.~Raghavan, S.~Rajagopalan, R.~Stata,
  A.~Tomkins, and J.~Wiener.
\newblock Graph structure in the web.
\newblock {\em Comput. Netw.}, 33:309--320, June 2000.

\bibitem{dhillon}
K.~B. Dhillon~I.S., Yuqiang~Guan.
\newblock Weighted graph cuts without eigenvectors a multilevel approach.
\newblock {\em IEEE Transactions on Pattern Analysis and Machine Intelligence},
  29:1944--1957, November 2007.

\bibitem{dunn}
J.~C. Dunn.
\newblock {A Fuzzy Relative of the ISODATA Process and Its Use in Detecting
  Compact Well-Separated Clusters}.
\newblock {\em Journal of Cybernetics}, 3(3):32--57, 1973.

\bibitem{Faloutsos:1999:PRI:316194.316229}
M.~Faloutsos, P.~Faloutsos, and C.~Faloutsos.
\newblock On power-law relationships of the internet topology.
\newblock {\em SIGCOMM Comput. Commun. Rev.}, 29:251--262, August 1999.

\bibitem{Flake:2002:SIW:619073.621934}
G.~W. Flake, S.~Lawrence, C.~L. Giles, and F.~M. Coetzee.
\newblock Self-organization and identification of web communities.
\newblock {\em Computer}, 35:66--71, March 2002.

\bibitem{newman}
M.~Girvan and M.~Newman.
\newblock Community structure in social and biological networks.
\newblock {\em Proceedings of the National Academy of Sciences of the United
  States of America}, 19:7821--7826, 2001.

\bibitem{erez}
E.~Hartuv and R.~Shamir.
\newblock A clustering algorithm based on graph connectivity.
\newblock {\em Information Processing Letter}, 76(4-6):175--181, 2000.

\bibitem{Hendrickson:1995:ISG:203046.203060}
B.~Hendrickson and R.~Leland.
\newblock An improved spectral graph partitioning algorithm for mapping
  parallel computations.
\newblock {\em SIAM J. Sci. Comput.}, 16:452--469, March 1995.

\bibitem{Karypis:1997:MHP:266021.266273}
G.~Karypis, R.~Aggarwal, V.~Kumar, and S.~Shekhar.
\newblock Multilevel hypergraph partitioning: application in vlsi domain.
\newblock In {\em Proceedings of the 34th annual Design Automation Conference},
  DAC '97, pages 526--529, New York, NY, USA, 1997. ACM.

\bibitem{Karypis:1998:FHQ:305219.305248}
G.~Karypis and V.~Kumar.
\newblock A fast and high quality multilevel scheme for partitioning irregular
  graphs.
\newblock {\em SIAM J. Sci. Comput.}, 20:359--392, December 1998.

\bibitem{kernighan}
L.~S. Kernighan B.~W.
\newblock An efficient heuristic procedure for partitioning graphs.
\newblock {\em Bell Systems Technical Journal}, 49:291--307, 1970.

\bibitem{Khandekar:2009:GPU:1538902.1538903}
R.~Khandekar, S.~Rao, and U.~Vazirani.
\newblock Graph partitioning using single commodity flows.
\newblock {\em J. ACM}, 56:19:1--19:15, July 2009.

\bibitem{csermely}
I.~Kovács, R.~Palotai, M.~Szalay, and P.~Csermely.
\newblock {Community landscapes: a novel, integrative approach for the
  determination of overlapping network modules.}
\newblock {\em PLoS ONE 5, e12528}, 2010.

\bibitem{DBLP:conf/tamc/2010}
J.~Kratochv\'{\i}l, A.~Li, J.~Fiala, and P.~Kolman, editors.
\newblock {\em Theory and Applications of Models of Computation, 7th Annual
  Conference, TAMC 2010, Prague, Czech Republic, June 7-11, 2010. Proceedings},
  volume 6108 of {\em Lecture Notes in Computer Science}. Springer, 2010.

\bibitem{lancichinetti-2009-11}
A.~Lancichinetti, S.~Fortunato, and J.~Kertesz.
\newblock Detecting the overlapping and hierarchical community structure of
  complex networks.
\newblock {\em NEW JOURNAL OF PHYSICS}, 11:033015, 2009.

\bibitem{Lang:1993:FNC:313559.313755}
K.~Lang and S.~Rao.
\newblock Finding near-optimal cuts: an empirical evaluation.
\newblock In {\em Proceedings of the fourth annual ACM-SIAM Symposium on
  Discrete algorithms}, SODA '93, pages 212--221, Philadelphia, PA, USA, 1993.
  Society for Industrial and Applied Mathematics.

\bibitem{DBLP:conf/nips/2000}
T.~K. Leen, T.~G. Dietterich, and V.~Tresp, editors.
\newblock {\em Advances in Neural Information Processing Systems 13, Papers
  from Neural Information Processing Systems (NIPS) 2000, Denver, CO, USA}. MIT
  Press, 2001.

\bibitem{Leighton:1999:MMM:331524.331526}
T.~Leighton and S.~Rao.
\newblock Multicommodity max-flow min-cut theorems and their use in designing
  approximation algorithms.
\newblock {\em J. ACM}, 46:787--832, November 1999.

\bibitem{Leskovec:2008:PVL:1367497.1367620}
J.~Leskovec and E.~Horvitz.
\newblock Planetary-scale views on a large instant-messaging network.
\newblock In {\em Proceeding of the 17th international conference on World Wide
  Web}, WWW '08, pages 915--924, New York, NY, USA, 2008. ACM.

\bibitem{DBLP:journals/rsa/LovaszS93}
L.~Lov{\'a}sz and M.~Simonovits.
\newblock Random walks in a convex body and an improved volume algorithm.
\newblock {\em Random Struct. Algorithms}, 4(4):359--412, 1993.

\bibitem{Meila01arandom}
M.~Meila and J.~Shi.
\newblock A random walks view of spectral segmentation.
\newblock In {\em Proceedings of the Eighth International Workshop on
  Artificial Intelligence and Statistics}, San Francisco, CA, USA, 2001. Morgan
  Kauffman.

\bibitem{mendes}
J.~Mendes and S.~Dorogovtsev.
\newblock Evolution of networks : From biological nets to the internet.
\newblock {\em Advances in Physics}, 51:1079--1187, 2002.

\bibitem{mislove-2009-socialnetworksthesis}
A.~Mislove.
\newblock {\em Online Social Networks: Measurement, Analysis, and Applications
  to Distributed Information Systems}.
\newblock PhD thesis, Rice University, Department of Computer Science, May
  2009.

\bibitem{mislove-2007-socialnetworks}
A.~Mislove, M.~Marcon, K.~P. Gummadi, P.~Druschel, and B.~Bhattacharjee.
\newblock {Measurement and Analysis of Online Social Networks}.
\newblock In {\em Proceedings of the 5th ACM/Usenix Internet Measurement
  Conference (IMC'07)}, San Diego, CA, October 2007.

\bibitem{newman2}
M.~Newman.
\newblock The structure and function of a complex networks.
\newblock {\em SIAM Review}, 45:167--256, 2003.

\bibitem{DBLP:conf/wea/2005}
S.~E. Nikoletseas, editor.
\newblock {\em Experimental and Efficient Algorithms, 4th
  InternationalWorkshop, WEA 2005, Santorini Island, Greece, May 10-13, 2005,
  Proceedings}, volume 3503 of {\em Lecture Notes in Computer Science}.
  Springer, 2005.

\bibitem{Orecchia:2008:PGV:1374376.1374442}
L.~Orecchia, L.~J. Schulman, U.~V. Vazirani, and N.~K. Vishnoi.
\newblock On partitioning graphs via single commodity flows.
\newblock In {\em Proceedings of the 40th annual ACM symposium on Theory of
  computing}, STOC '08, pages 461--470, New York, NY, USA, 2008. ACM.

\bibitem{DBLP:journals/corr/abs-0810-4061}
P.~Orponen, S.~E. Schaeffer, and V.~A. Gayt{\'a}n.
\newblock Locally computable approximations for spectral clustering and
  absorption times of random walks.
\newblock {\em CoRR}, abs/0810.4061, 2008.

\bibitem{palia}
G.~Palla, I.~Derenyi, I.~Farkas, and T.~Vicsek.
\newblock Uncovering the overlapping community structure of complex networks in
  nature and society.
\newblock {\em Nature}, 435:814--818, 2005.

\bibitem{Pothen:1990:PSM:84514.84521}
A.~Pothen, H.~D. Simon, and K.-P. Liou.
\newblock Partitioning sparse matrices with eigenvectors of graphs.
\newblock {\em SIAM J. Matrix Anal. Appl.}, 11:430--452, May 1990.

\bibitem{jruan}
J.~Ruan and W.~Zhang.
\newblock Identifying hierarchical community structures in biological networks.
\newblock {\em Algorithmic Biology}, 2006.

\bibitem{dongen}
S.~van Dongen.
\newblock {\em Graph clustering by flow simulation}.
\newblock PhD thesis, Universiteit Utrecht, Utrecht, The Netherlands, 2000.

\bibitem{wagner}
A.~Wagner and D.~Fell.
\newblock The small world inside large metabolic networks.
\newblock {\em Proc Biol Sci.}, 268:1803--1810, 2001.

\bibitem{PhysRevLett.86.2050}
F.~Wang and D.~P. Landau.
\newblock Efficient, multiple-range random walk algorithm to calculate the
  density of states.
\newblock {\em Phys. Rev. Lett.}, 86:2050--2053, Mar 2001.

\bibitem{DBLP:conf/sofsem/2006}
J.~Wiedermann, G.~Tel, J.~Pokorn{\'y}, M.~Bielikov{\'a}, and J.~Stuller,
  editors.
\newblock {\em SOFSEM 2006: Theory and Practice of Computer Science, 32nd
  Conference on Current Trends in Theory and Practice of Computer Science,
  Mer\'{\i}n, Czech Republic, January 21-27, 2006, Proceedings}, volume 3831 of
  {\em Lecture Notes in Computer Science}. Springer, 2006.

\bibitem{zachary}
W.~Zachary.
\newblock An information flow model for conflict and fission in small groups.
\newblock {\em Journal of Anthropological Research}, 33:452--473, 1977.

\bibitem{PhysRevE.78.046705}
C.~Zhou and J.~Su.
\newblock Optimal modification factor and convergence of the wang-landau
  algorithm.
\newblock {\em Phys. Rev. E}, 78:046705, Oct 2008.

\end{thebibliography}
\end{document}